# Black-box Integration of Heterogeneous Modeling Languages for Cyber-Physical Systems

Markus Look, Antonio Navarro Perez, Jan Oliver Ringert, Bernhard Rumpe, and Andreas Wortmann

Software Engineering
RWTH Aachen University
http://www.se-rwth.de/

**Abstract.** Robots belong to a class of Cyber-Physical Systems where complex software as a mobile device has to fulfill tasks in a complex environment. Modeling robotics applications for analysis and code generation requires modeling languages for the logical software architecture and the system behavior. The MontiArcAutomaton modeling framework integrates six independently developed modeling languages to model robotics applications: a component & connector architecture description language, automata, I/O tables, class diagrams, OCL, and a Java DSL. We describe how we integrated these languages into MontiArcAutomaton a-posteriori in a black-box integration fashion.

## 1 Introduction

Robots are a class of Cyber-Physical Systems (CPS) [5] where complex software has to fulfill tasks in a continuous environment. A successful robotics application has to consider different aspects of hardware and software: ranging from spatial properties of the robot to kinematics and hardware drivers to the software architecture and the behavior of the system. MontiArcAutomaton [8,9] integrates six independently developed, heterogeneous modeling languages:

- the MontiArc architecture description language [2] to model the software architecture of the system as (atomic and decomposed) components, connected with unidirectional connectors via their typed ports;
- I/O$^\omega$ automata [11,7] and I/O tables [10,14] to model the behavior of atomic components;
- UML/P class diagrams (CD) [12,13] model data types of ports and variables;
- the UML/P Object Constraint Language (OCL) [12,13] and a Java Modeling Language (Java ML), to model guards on I/O$^\omega$ automata transitions.

MontiArc, I/O$^\omega$ automata, I/O tables, CD, OCL, the Java ML, and MontiArcAutomaton are implemented as MontiCore [4] languages in form of context-free grammars with context condition checks. MontiCore facilitates language integration with a compositional symboltable framework – the Extensible Type System (ETS) [15] – which provides mechanisms for language aggregation (the MontiArcAutomaton language family aggregates class diagrams), language inheritance (MontiArcAutomaton extends MontiArc), and language embedding (I/O$^\omega$ automata embed Java and OCL). Crucially, this framework supports an a-posteriori integration of languages in a black-box integration fashion.

In this short paper we give an overview of a specific case of language composition for robotics applications. Comparisons of MontiCore's language composition and integration mechanisms with related approaches are presented in [4,15].

Fig. 1 illustrates MontiArcAutomaton's language integration mechanisms on the model of a simple robot. The robot drives forward until it approaches an obstacle, it then drives backwards, rotates, and drives forward again. Decomposition and interfaces of the components of `BumperBot` are modeled using the concepts inherited from MontiArc. The behavior of the component `BumpControl` is modeled with an I/O$^\omega$ automaton, where two transitions are guarded with embedded OCL expressions. The behavior of the component `Timer` is modeled as an I/O table and the behaviors of components wrapping hardware are implemented in the target platform language. The types of component ports and automaton variables are modeled in a class diagram.

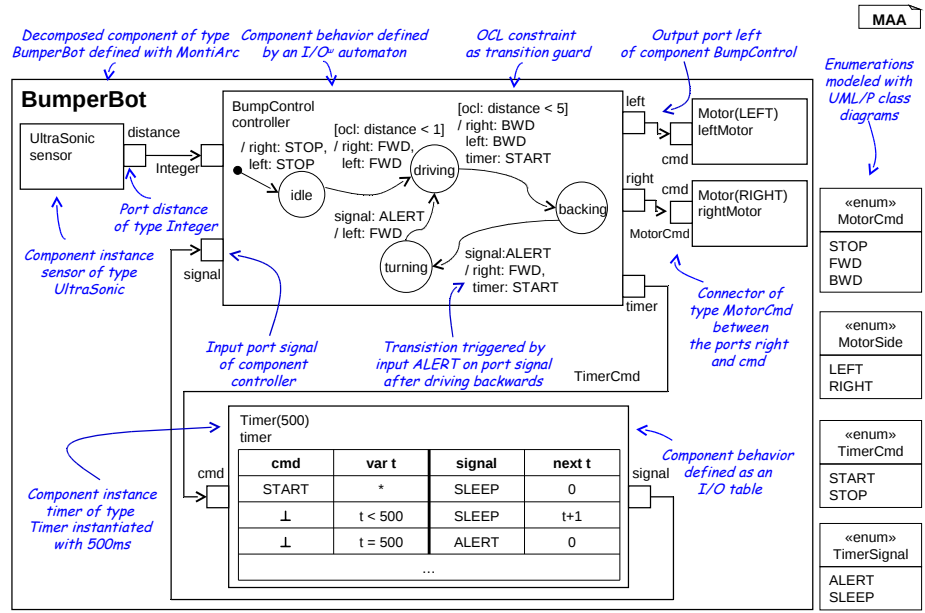

**Fig. 1.** The MontiArcAutomaton model of a robot integrating five modeling languages.

While Fig. 1 gives a graphical overview of the used models, the concrete, textual models used to model the CPS are provided in the ReMoDD repository[1].

---

[1] http://www.cs.colostate.edu/remodd/v1/content/black-box-integration-heterogeneous-modeling-languages-cps-supporting-materials

Within this paper and with the concrete models attached we show how the three mechanisms for integrating modeling languages are used within the MontiArcAutomaton modeling language to describe the multiple aspects of a robotics application. In the next sections we introduce the three language integration mechanisms used and show how these are realized in the MontiCore framework.

## 2 Language Integration Mechanisms

MontiCore offers three distinct mechanisms for integrating heterogeneous languages: embedding, aggregation, and inheritance. For a detailed explanation of these mechanisms see [4,15]. All mechanisms have different benefits in varying application scenarios, which will be presented in this section.

For MontiArcAutomaton we chose to model components and embedded automata in a single modeling artifact, as shown in Fig. 1. This is a tighter coupling than a definition in separate artifacts as, for instance, done when using UML component diagrams and UML class diagrams [6]. We believe that the syntactic integration helps to prevent inconsistencies when evolving either the structure of a component or its behavior [3].

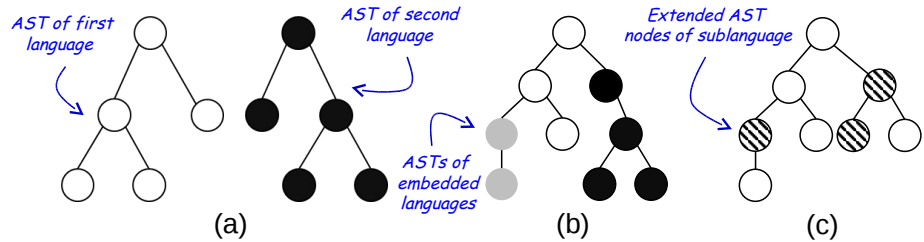

**Fig. 2.** The resulting abstract syntax trees (ASTs) when using (a) aggregation, (b) embedding, and (c) inheritance. Aggregation results in separate ASTs for each model. Embedding results in a single AST with subtrees embedded at the leaves of the host language. Inheritance results also in a single AST containing extended nodes of the sublanguage.

**Language Aggregation:** When modeling robotics applications, different orthogonal aspects, like the definition of a common type system as a class diagram and its usage, need to be expressed by an aggregated set of languages for specific purposes. Each aspect is modeled within a single artifact written in a specific language thus producing separate abstract syntax trees (ASTs) as shown in Fig. 2. This approach is used to aggregate the heterogeneous languages as shown in Fig. 3 and enables establishing a knowledge relationship between the models in terms of references.

**Language Embedding:** Language embedding offers using different languages within the same model. MontiCore allows the definition of external nonterminals within a grammar which allows for binding nonterminals of other grammars to these external nonterminals and thus embedding (parts of) the language. The AST in turn consists of objects belonging to the respective languages as shown in Fig. 2. This mechanism is especially useful when modeling different paradigms, e.g., embedding a language for modeling behavior in a language for modeling structure. We use this for embedding, e.g., I/O$^\omega$ automata in the MontiArcAutomaton language, as shown in Fig. 3.

**Language Inheritance:** Language inheritance can be utilized to refine or extend an existing language. MontiCore allows inheriting from existing nonterminals thus enabling extension of those languages as shown in Fig. 2. For the MontiArcAutomaton language we extended the language MontiArc by new nonterminals that allow the definition of embedded I/O$^\omega$ automata and variables as shown in Fig. 3. This approach is especially useful to reuse existing concepts of the super language while extending it by new concepts.

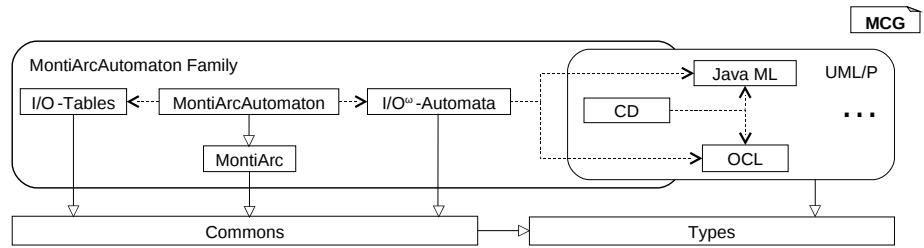

**Fig. 3.** An overview of the MontiArcAutomaton integration. Rounded corners show language aggregation. All other boxes show languages. Dashed arrows mark language embedding. Solid arrows mark language inheritance. The CD language is part of both the UML/P and the MontiArcAutomaton family aggregation.

## 3 Realization

In MontiCore, the integration of languages is implemented through the Extensible Type System (ETS) sub-framework. To that end, ETS provides various concepts and extension points that support an a-posteriori integration of aggregated, embedded and inherited languages in a black-box integration fashion.

The core elements of an ETS-based implementation are *namespaces*, *symboltables*, and *entries*. Their structure, construction, and integration are the key aspects of an implementation of integrated languages.

Entries are descriptions of model elements and implemented as subclasses of an abstract base class. For instance, `ComponentEntry` instances describe the essence of component type definitions (e.g., `BumperBot`, `BumpControl`). Entries can be nested into each other and reference each other even if the model elements they describe are not part of the same model. This reference mechanism is, in addition, subject to visibility rules that reflect the modeling language visibility policies. For instance, the interface of component types is visible to other models, whereas its internal structure is not. Thereby, the structure of entries resembles the essential structure of integrated models.

Namespaces, symboltables, resolvers, and entries are, among other modules not mentioned in this paper, custom implemented by a developer for single modeling languages via extension points in the ETS framework [15]. Crucially, these implementations are encapsulated modules that can be integrated with each other without a need for invasive modifications. To this end, the ETS provides several means that allow an a-posteriori integration of modeling languages, based on any combination of language aggregation, embedding, and inheritance.

A core mechanism is the adaptation of entries from different languages by entry adapters. Entry adapters are implementations of the adapter pattern [1]. An entry adapter translates the interface of an entry from one language such that it corresponds to an interface of an entry from another language. Their common use case is the translation of a modeled concept between a language that defines the concept and a language that references it. For instance, port declarations in components, described by PortEntry, reference a type definition, described by ArcdTypeEntry, which in our case originates in class diagram models where types are described by CDTypeEntry. In the implementation of the integration of these two languages, CDTypeEntry2ArcdTypeAdapter instances adapt class type entries such that they conform to port type entries. Fig. 4 illustrates this scenario. In this way, modeling concepts from different languages are integrated. The ETS framework applies adapters automatically.

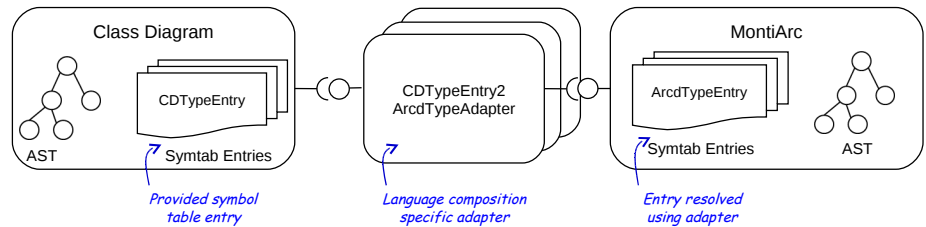

**Fig. 4.** Example of the integration of MontiArc and CDs by adapting CD types to the MontiArc symbol table.

Beyond this basic infrastructure, MontiCore provides the developer with various tools and extension points for the implementation of analyses and syntheses on heterogeneous models. In general, such custom modules are all based on an implementation of the visitor pattern [1]. These visitors can traverse language-specific parts of heterogeneous ASTs. Furthermore, they can be combined into composite visitors for specific compositions of languages. Visitors are, among other use cases, employed for context conditions, like type checks, and model transformations as well as for syntheses like code generation. For instance, a composite code generator executes different code generation components for components and I/O$^\omega$ automata while traversing the models of the aggregated component and automaton languages.

Visitor executions are organized into workflows. Workflows are modules provided by the developer that are registered for specific (aggregated) modeling languages. They are automatically executed for every model of that language. Workflows can refer to information from other models (potentially models of other languages) by resolving their entries through the ETS framework.

## 4  Conclusion

Integration of modeling languages facilitates the efficient development of complex software systems. We have illustrated how language aggregation, language embedding and language inheritance can be used to model both architecture and behavior of CPS in integrated models. The MontiCore framework supports the application of these mechanisms in a black-box fashion through its ETS framework and its workflows and visitors components.